\documentclass[a4paper]{jpconf}
\usepackage{graphicx}
\begin{document}
\title{Anisotropic flow of identified particles in Pb--Pb collisions at $\sqrt{s_{_{\rm NN}}} =$  2.76 TeV with the ALICE detector}

\author{You Zhou (for the ALICE Collaboration)}

\address{Nikhef, Science Park 105, 1098 XG Amsterdam, The Netherlands}
\address{Utrecht University, P.O.Box 80000, 3508 TA Utrecht, The Netherlands}

\ead{you.zhou@cern.ch}

\begin{abstract}
The measurement of anisotropic flow of identified particles, including charged pion, kaon, (anti-)proton as well as strange and multi-strange particles (${\rm K_{S}^{0}}$, ${\rm \Lambda}$, ${\rm \Xi}$ and ${\rm \Omega}$) are measured in Pb--Pb collisions at $\sqrt{s_{_{{\rm NN}}} }$ = 2.76 TeV with the ALICE detector. The results are compared to measurements at RHIC and hydrodynamic model calculations. The mass splitting and the scaling with the number of constituent quarks of the anisotropic flow are also discussed. 
\end{abstract}

\section{Introduction}

Anisotropic flow~\cite{Ollitrault:1992bk}, described by the coefficients in the Fourier expansion of the azimuthal particle distributions with respect to the symmetry planes, 
is an important observable to study the properties of the hot and dense matter, the
Quark Gluon Plasma (QGP), created in heavy-ion collisions~\cite{Voloshin:2008dg}.
The anisotropic flow of identified particles allows us to probe the freeze-out conditions of the system, 
{\it e.g.} the freeze-out temperature, and the radial flow. It was measured at RHIC~\cite{Adams:2003am, Adare:2006ti} and could be successfully described by hydrodynamic calculations~\cite{Heinz:2011kt}. 
Moreover, it is very sensitive to the partonic degrees of freedom at the early time of the heavy-ion 
collisions. Therefore, the identified particle elliptic flow, $v_{2}$, and triangular flow, $v_{3}$, measured at the LHC is expected to play a pivotal role in constraining the initial conditions as well as the shear viscosity over entropy density ($\eta/s$) in the hydrodynamic calculations.

We report on $v_{2}$ and $v_{3}$ of identified particles measured in $\sqrt{s_{_{\rm NN}}} =$ 2.76 TeV Pb--Pb collisions by the ALICE Collaboration.  The measurements are compared to those at top RHIC energy and to ideal hydrodynamic calculations~\cite{Heinz:2011kt}. The scaling with number of constituent quarks (2 for mesons and 3 for baryons) is also tested for both $v_{2}$ and $v_{3}$.

\section{Data Analysis}
The data sample collected by ALICE in the first Pb--Pb run at the Large Hadron Collider was used in this analysis. About 16 million Pb--Pb events were recorded with a minimum-bias trigger, based on signals from two VZERO (-3.7$\textless \eta \textless$-1.7 and 2.8$\textless \eta \textless$5.1) and on the Silicon Pixel Detector. The VZERO was also used for the determination of the collision centrality and of the symmetry planes. Charged particles are reconstructed using the Inner Tracking System and the Time Projection Chamber (TPC) with full azimuthal coverage for pseudo-rapidity range $|\eta|\textless$0.9. The particle identification was performed using the specific ionisation energy loss in the TPC and time-of-flight measurements in the Time-Of-Flight detector. The strange and multi-strange particles were reconstructed via their weak decay channels, which are: $\rm {K_{S}^{0} \rightarrow \pi^{+} \pi^{-}}$, ${\rm \Lambda \rightarrow p\pi^{-}~(\overline{\Lambda} \rightarrow \overline{p}\pi^{+})}$
, ${\rm \Xi^{-} \rightarrow \Lambda \pi^{-}~(\overline{\Xi}^{+} \rightarrow \overline{\Lambda} \pi^{+})}$ and ${\rm \Omega^{-}\rightarrow \Lambda K^{-} ~(\overline{\Omega}^{+}\rightarrow \overline{\Lambda} K^{+})}$. In order to suppress combinatorial background, additional topological cuts have been applied. Anisotropic flow was measured with the scalar product~\cite{Adler:2002pu} and event plane~\cite{Poskanzer:1998yz} methods with large separation  in pseudo-rapidity between the reference particles and the particle of interest ($|\Delta \eta|\textgreater1$ for SP method and $|\Delta \eta|\textgreater2$ for EP method), to reduce the correlations not associated with the symmetry plane, the so called non-flow.

\section{Results}

\begin{figure}[h]
\includegraphics[width=0.49\textwidth]{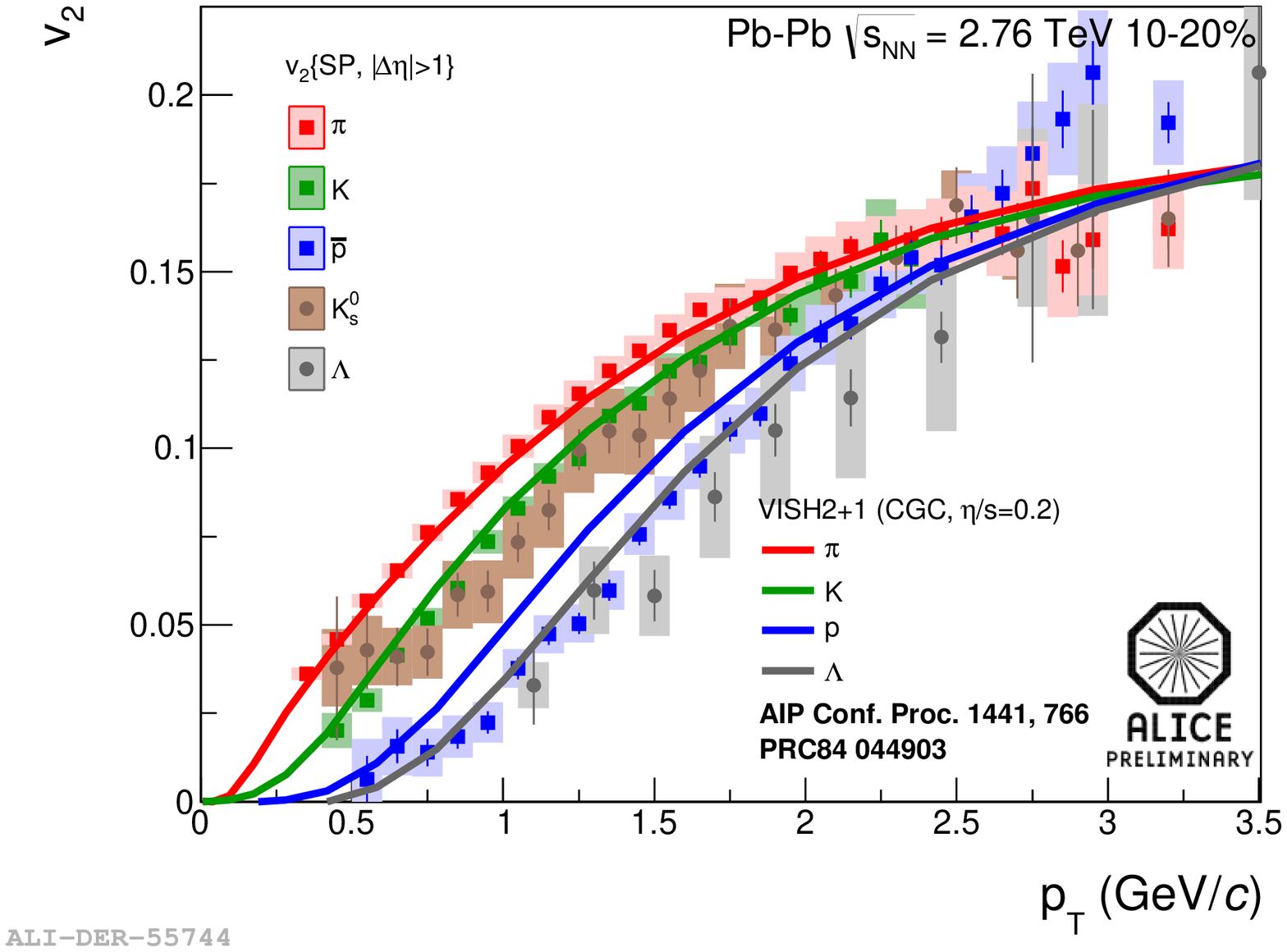}
\includegraphics[width=0.49\textwidth]{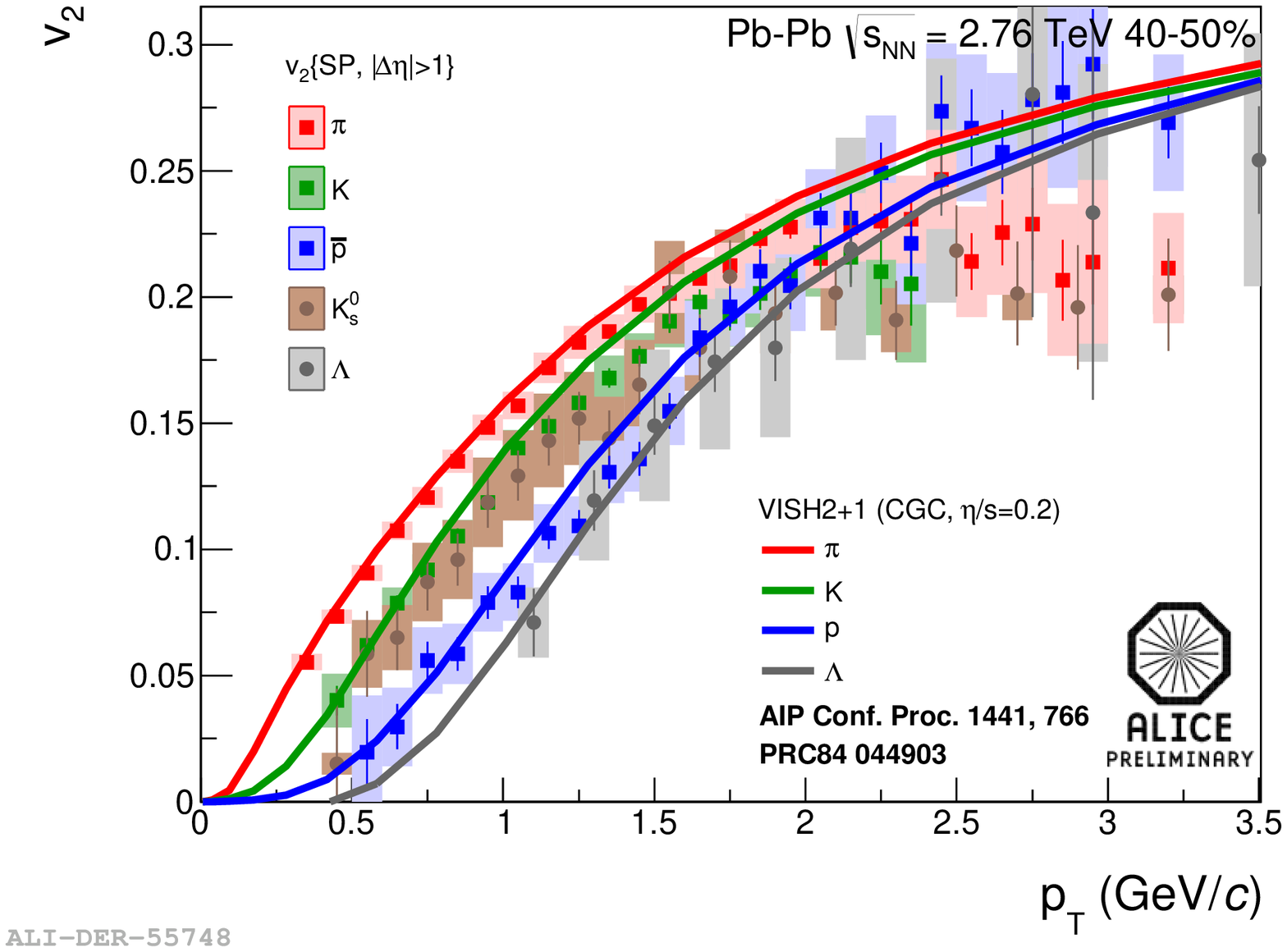}
\caption{\label{label} Identified particle $v_{2}(p_{\rm T})$ measured by ALICE for 10--20$\%$ (left) and 40--50$\%$ (right) centrality classes, together with viscous hydrodynamic model calculations~\cite{Heinz:2011kt}. }
\end{figure}

\begin{figure}[h]
\includegraphics[width=0.49\textwidth]{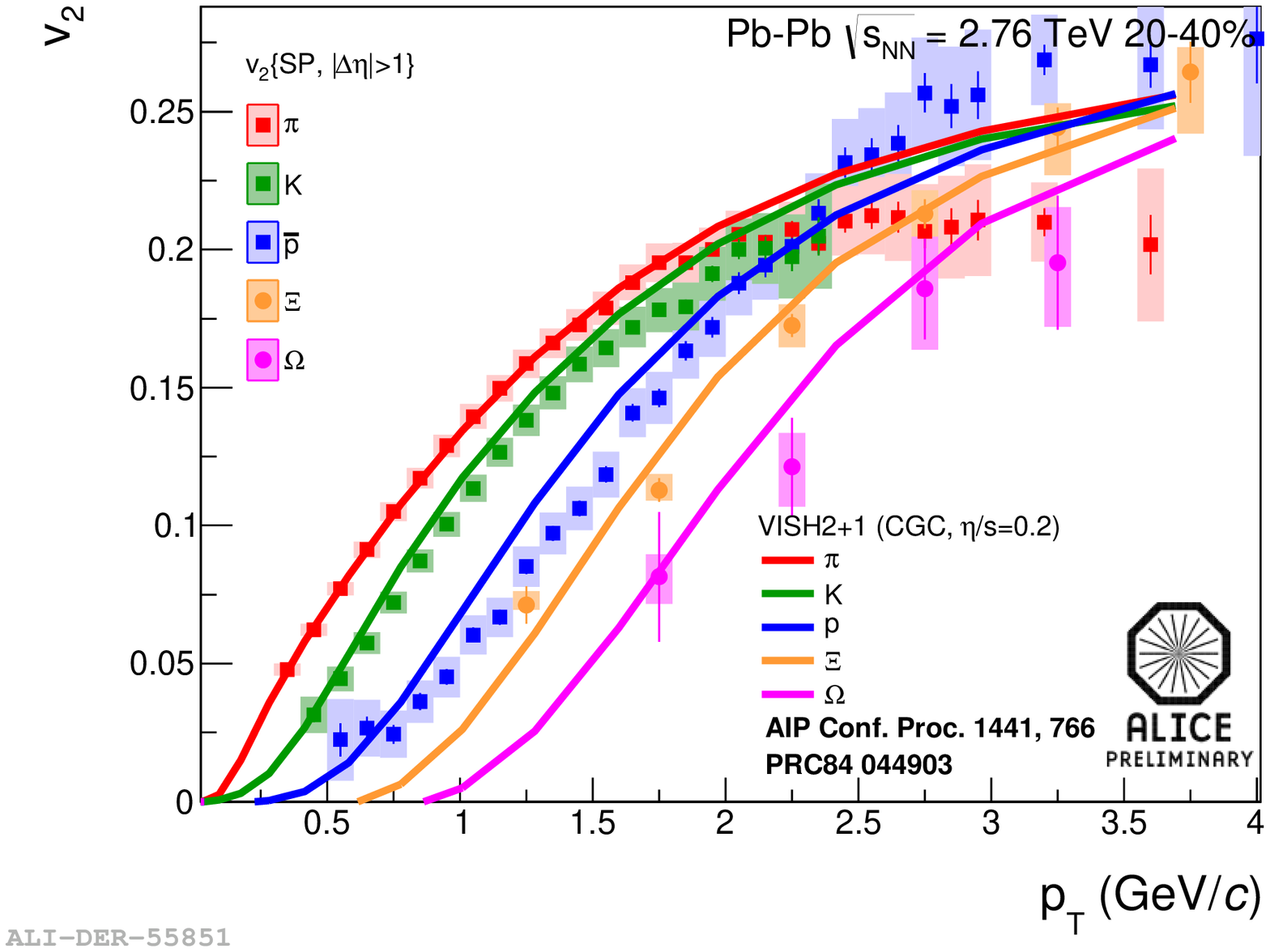}
\includegraphics[width=0.49\textwidth]{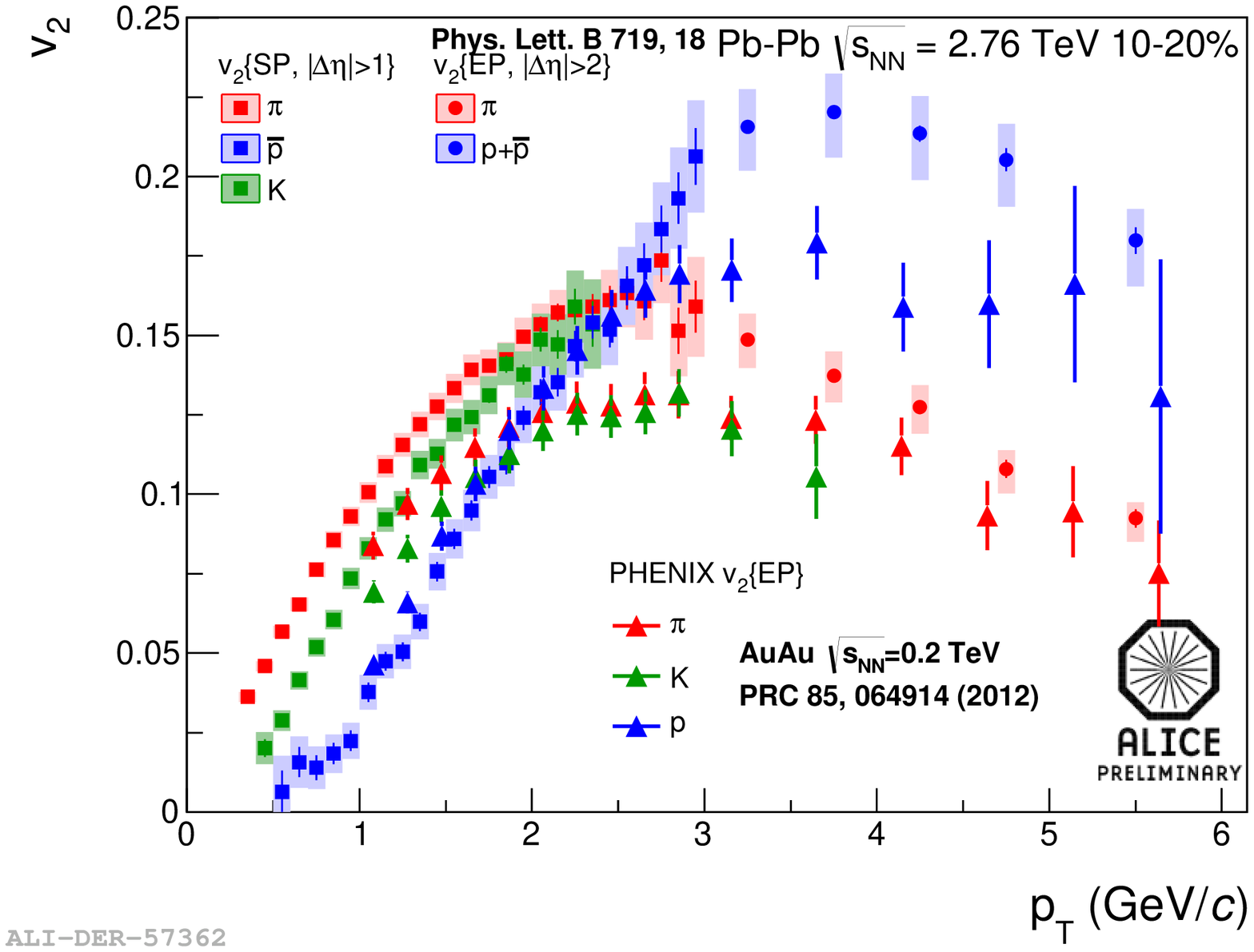}
\caption{\label{label} $\Xi$ and $\Omega$ $v_{2}(p_{\rm T})$ measured by ALICE in 20--40 $\%$ centrality class (left) compared to hydrodynamic calculations~\cite{Heinz:2011kt}; pion, kaon, proton $v_{2}(p_{\rm T})$ measured by ALICE compared to PHENIX~\cite{Adare:2012vq} in 10--20$\%$ centrality (right).}
\end{figure}

Figure 1 shows the elliptic flow for ${\rm \pi}$, ${\rm K}$, ${\rm p(\overline{p})}$, ${\rm K_{S}^{0}}$, $\Lambda(\overline{\Lambda})$ as a function of $p_{\rm T}$ in different centrality classes along with hydrodynamical calculations~\cite{Heinz:2011kt}. A clear mass splitting is seen in the presented $p_{\rm T}$ region, more pronounced in central collisions, indicating a stronger radial flow built there.
Viscous hydrodynamic model calculations (VISH2+1), using Color Glass Condensate initial condition with $\eta/s$ = 0.20, quantitatively reproduce the mass splitting of $v_{2}$ for most of the particles, while a better agreement is observed in peripheral collisions. In central collisions, the hydrodynamic calculations overestimate the proton $v_{2}$,  indicating that proton might freeze-out later (with larger radial flow) compared to pion and kaon. It also suggests an important role of hadronic interactions in reproducing proton $v_{2}$.
A similar mass splitting can be observed for the $v_{2}$ of $\Xi$ and $\Omega$ shown in Fig. 2 (left).

Figure 2 (right) presents a comparison of the $p_{\rm T}$-differential $v_{2}$ measured in Au--Au collisions at $\sqrt{s_{_{\rm NN}}} =$  200 GeV from PHENIX~\cite{Adare:2012vq} and ALICE for different particles species. It is seen that the $v_{2}$ of ${\rm \pi}$ and K at the LHC is slightly higher compared to RHIC.
However, the $v_{2}$ of proton is lower at low $p_{\rm T}$ but higher at high $p_{\rm T}$ compared to the one at RHIC, reflecting the effect of larger radial flow produced at LHC, as expected from hydrodynamical model predictions~\cite{Heinz:2011kt}.

\begin{figure}[h]
\includegraphics[width=0.49\textwidth]{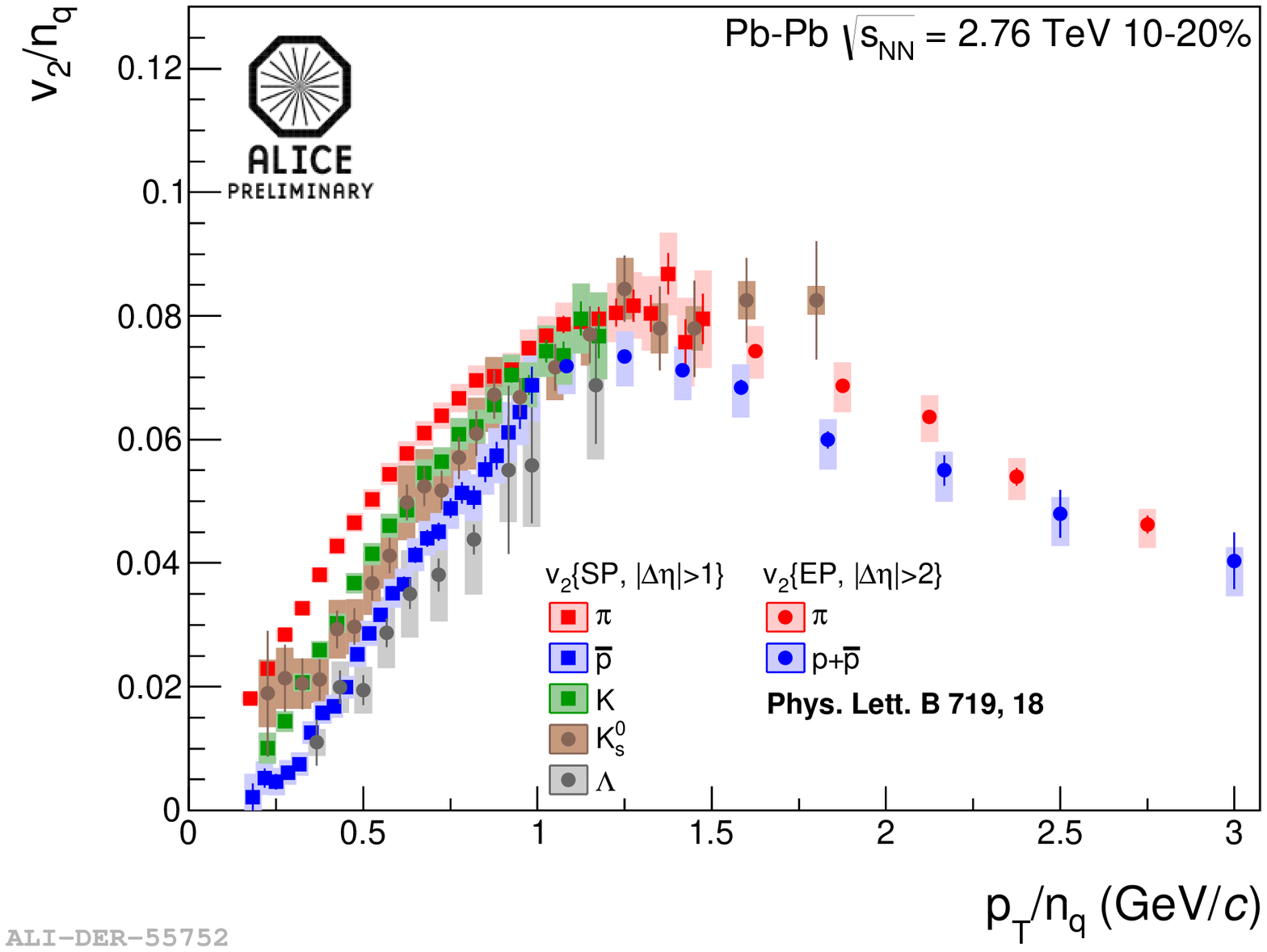}
\includegraphics[width=0.49\textwidth]{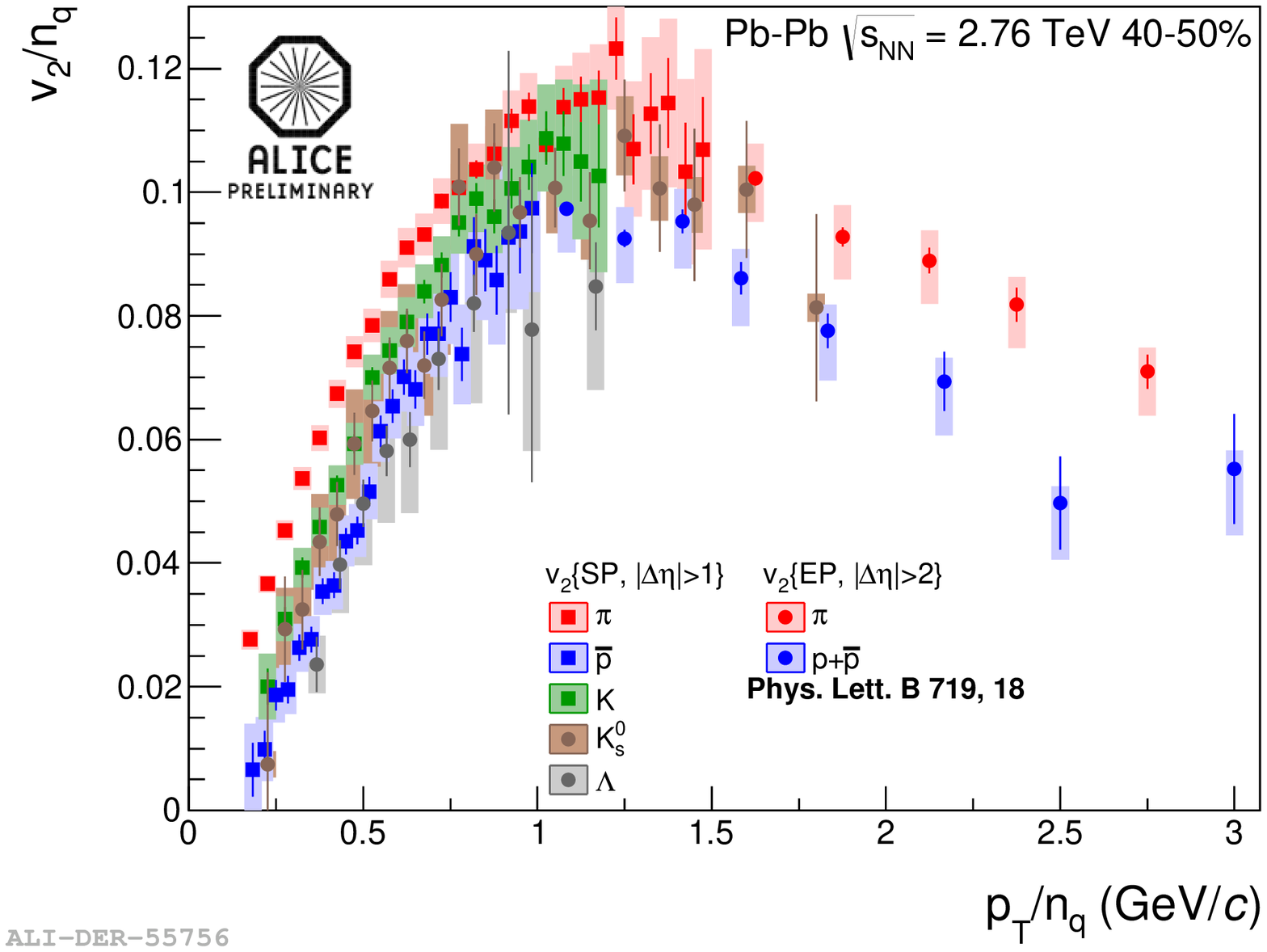}
\caption{\label{label} Identified particle $v_{2}(p_{\rm T})$ scaling with $n_{q}$ as a function of $p_{\rm T}/n_{q}$ in 10--20$\%$ (left) and 40--50$\%$ (right) centrality classes.}
\end{figure}

\begin{figure}[h]
\includegraphics[width=0.49\textwidth]{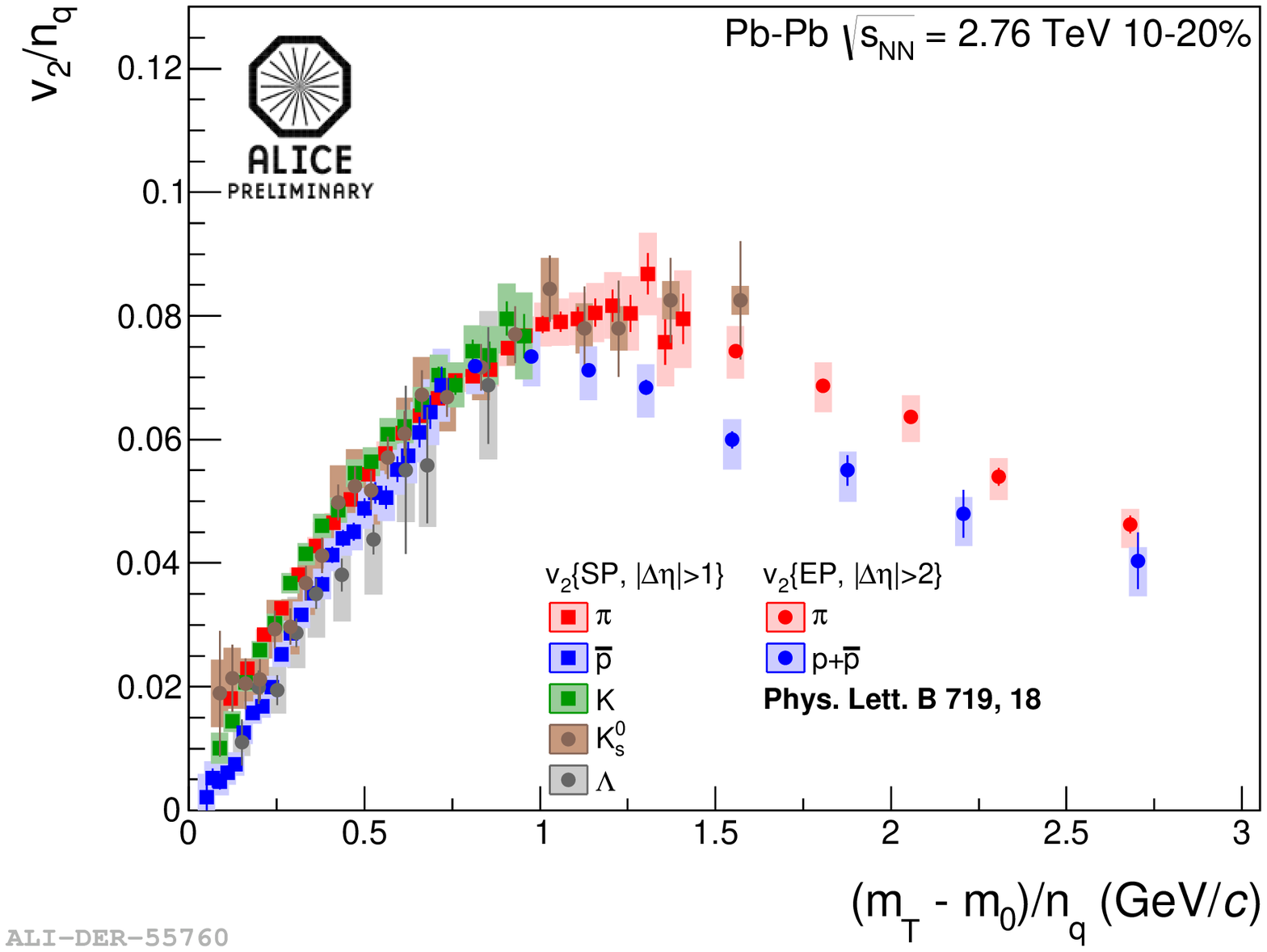}
\includegraphics[width=0.49\textwidth]{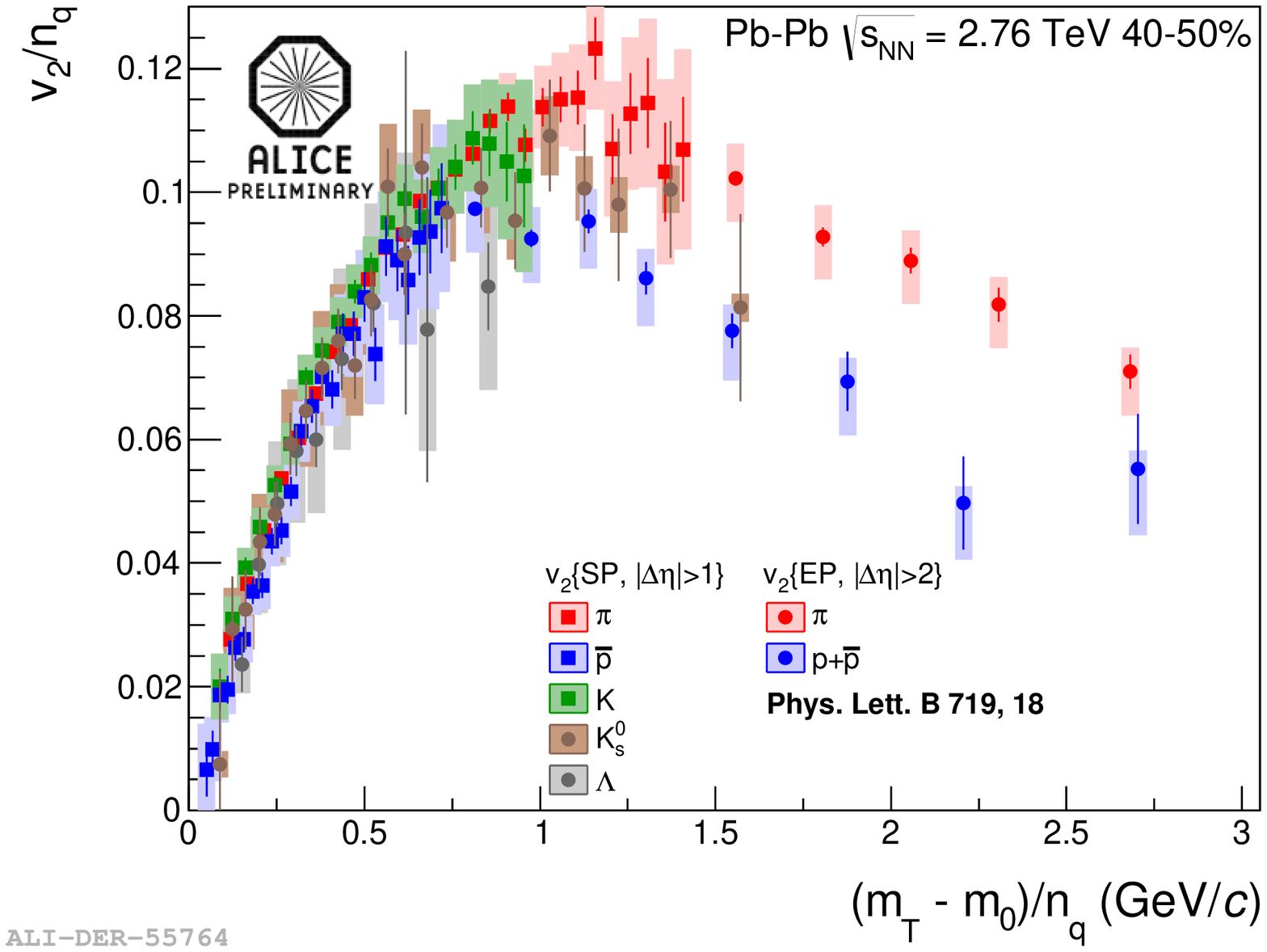}
\caption{\label{label}Identified particle $v_{2}(p_{\rm T})$ scaling with $n_{q}$ as a function of $KE_{\rm T}/n_{q}$ in 10--20$\%$ (left) and 40--50$\%$ (right) centrality classes.}
\end{figure}

In addition, the scaling with the number of constituent quarks observed at RHIC was used to support the picture that the collectivity already develops at the partonic level in heavy-ion collisions~\cite{Adams:2003am,  Adare:2006ti}. It can also serve as a test of hadron production via quark coalescence mechanism.
Figures 3 and 4 show the elliptic flow of identified particles scaled by $n_{q}$  as a function of the transverse momentum and transverse kinetic energy ($KE_{\rm T} = \sqrt{m_{0}^{2} + p_{\rm T}^{2}} - m_{0}$) per $n_{q}$, respectively. 
It appears that the scaling with $n_{q}$ works only approximately at the LHC energy, within 20$\%$ at $p_{\rm T}/n_{q} \sim$ 1.2 GeV/${\it c}$.
For the $KE_{\rm T}$ scaling, the $v_{2}/n_{q}$ of proton is clearly lower than that of pion for $KE_{\rm T}/n_{q}$ above 1 GeV/${\it c}$. The scaling is violated at LHC, in contrast to the observation at top RHIC energy.

\begin{figure}[h]
\includegraphics[width=0.46\textwidth]{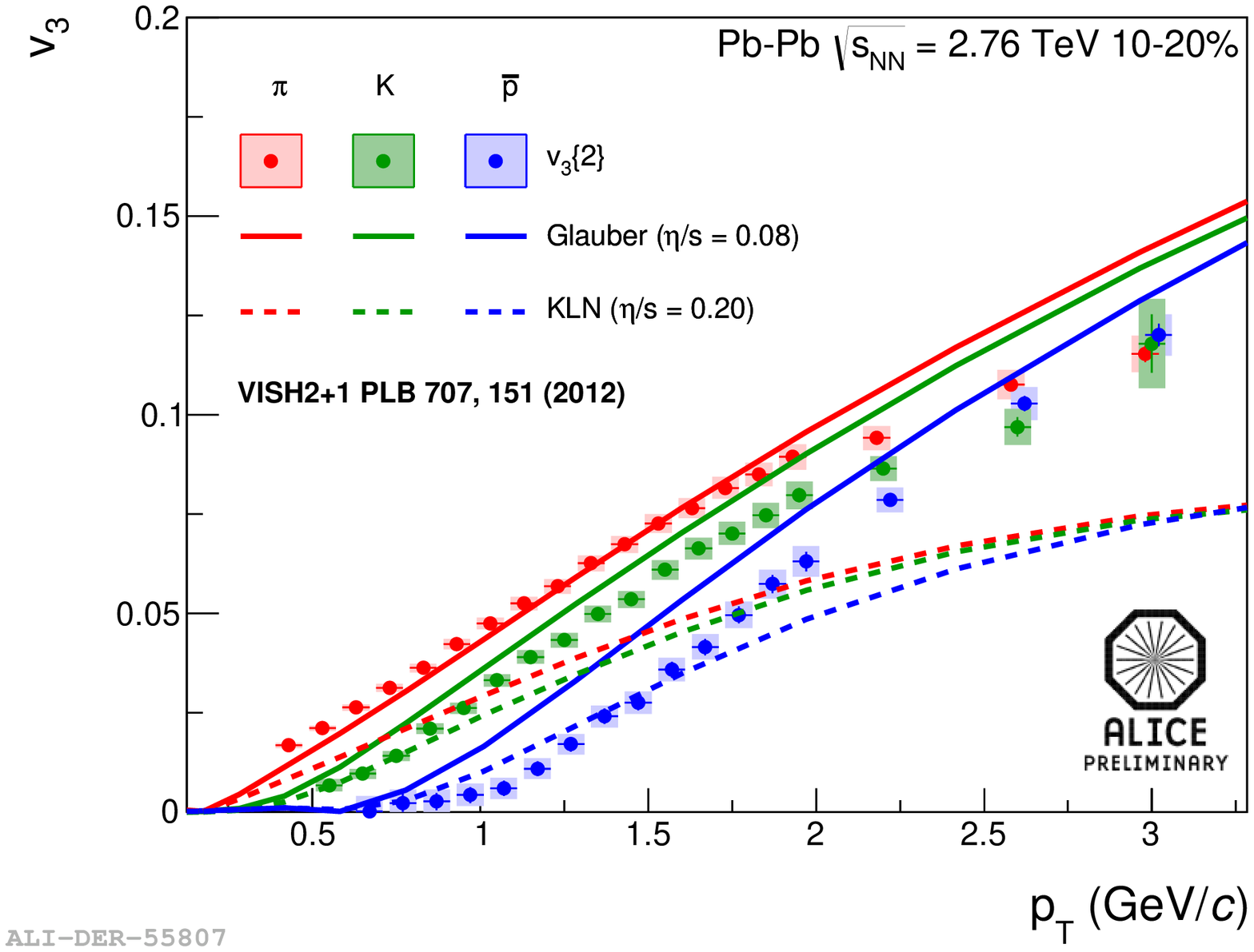}
\includegraphics[width=0.49\textwidth]{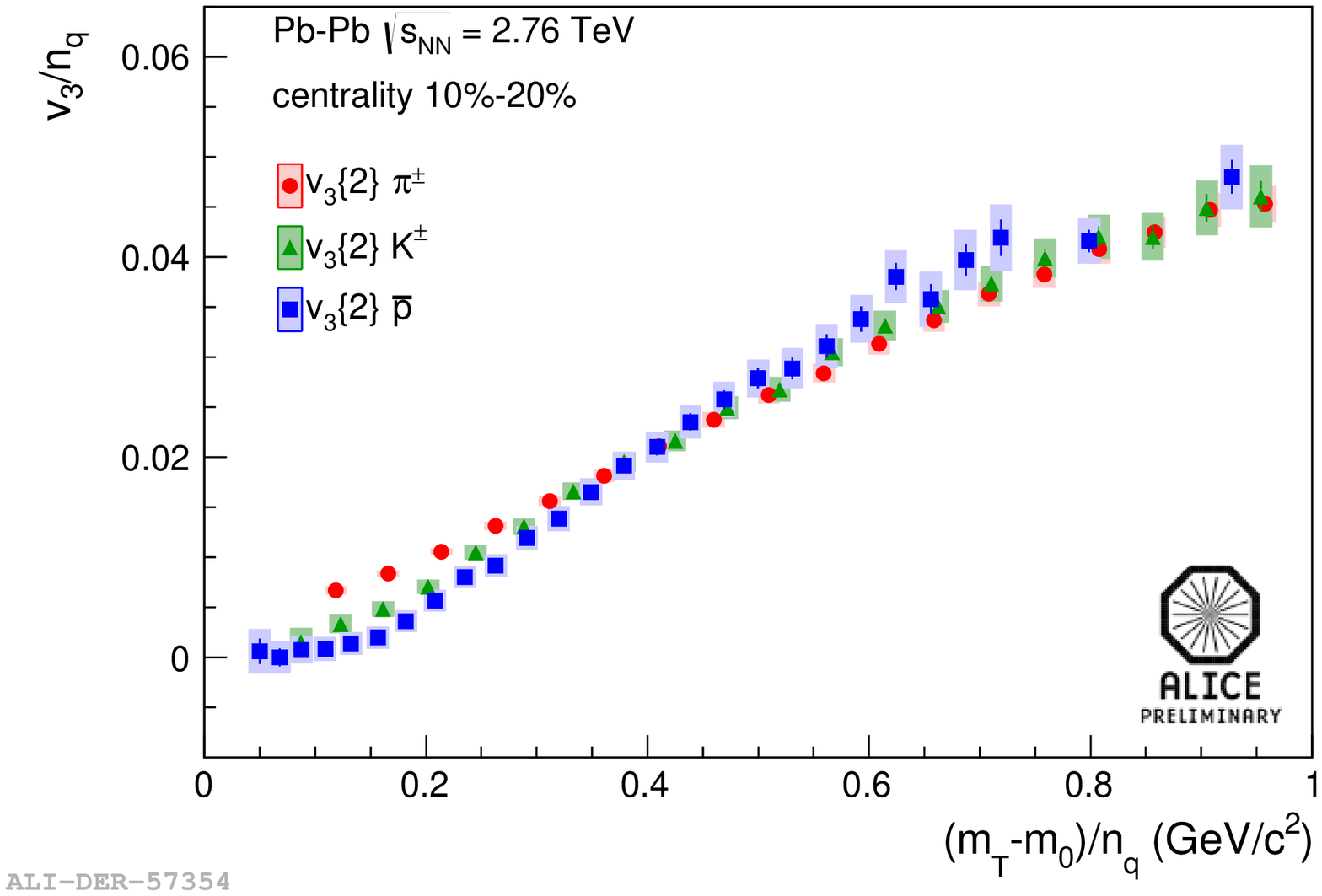}
\caption{\label{label}Identified particle $v_{3}(p_{\rm T})$ (left) and the $KE_{\rm T}/n_{q}$ scaling in 10--20$\%$ centrality.}
\end{figure}

Figure 5 (left) presents the triangular flow $v_{3}$ of identified particle as a function of transverse momentum in 10--20$\%$ centrality.
The $v_{3}$ exhibits similar features as the $v_{2}$: a clear mass splitting is observed at low $p_{\rm T}$ as predicted from hydrodynamic calculations while the protons cross pions at intermediate $p_{\rm T}$ as expected from the quark coalescence mechanism. 
Hydrodynamic calculations can not quantitatively reproduce the $v_{3}$ results, neither using CGC initial condition with $\eta/s$ = 0.20 nor with Glauber initial condition with $\eta/s$ = 0.08.
The $KE_{\rm T}$ scaling for $v_{3}$ works better than for $v_{2}$, but it is still only approximate as can be seen in Fig. 5 (right).

\section{Summary}
We presented the anisotropic flow measurements of identified particles in Pb--Pb collisions at $\sqrt{s_{_{\rm NN}}} =$  2.76 TeV with the ALICE detector. 
A good agreement with hydrodynamic model calculations is found for most of the particles. Compared to the measurements at top RHIC energy, the results suggest a larger radial flow at LHC, as predicted by hydrodynamic calculations. The scaling with the number of constituent quarks in only approximate within 20$\%$ at $p_{\rm T}/n_{q}\sim$ 1.2 GeV/${\it c}$. The $KE_{\rm T}/n_{q}$ scaling is broken over the all $p_{\rm T}$ range at the LHC energy. The $v_{3}$ shows similar features to $v_{2}$: a mass splitting at low $p_{\rm T}$, while protons cross pions at intermediate $p_{\rm T}$. The $KE_{\rm T}/n_{q}$ scaling works better for $v_{3}$ but it is still only approximate.

\smallskip

\end{document}